\documentclass{tMOP2e}

\begin{document}

\doi{10.1080/0950034YYxxxxxxxx}
 \issn{1362-3044}
\issnp{0950-0340} \jvol{00} \jnum{00} \jyear{2008} \jmonth{10 January}
\markboth{Andersson, Barnett, Hillery and Oi}{Journal of Modern Optics}

\title{Decision problems with quantum black boxes}
\author{Mark Hillery$^{a}$,
Erika Andersson$^{b}$ $^{\ast}$\thanks{$^\ast$Corresponding author. Email: E.Andersson@hw.ac.uk\vspace{6pt}}, 
Stephen M. Barnett$^{c}$ and Daniel Oi$^{c}$\\
\vspace{6pt}  
$^{a}${\em{Department of Physics and Astronomy, Hunter College of CUNY, 695 Park Avenue,
New York, NY 10065 USA}};
$^{b}${\em{SUPA, Department of Physics, Heriot-Watt University, Edinburgh EH14 4AS,
UK}};
$^{c}${\em{SUPA, Department of Physics, University of Strathclyde,
Glasgow G4 0NG, UK}} \\
\vspace{6pt}
\received{\today} }
\maketitle

\begin{abstract}

We examine how to distinguish between unitary operators, when the exact form of the possible operators is not known. Instead we are supplied with ``programs" in the form of unitary transforms, which can be used as references for identifying the unknown unitary transform. All unitary transforms should be used as few times as possible. This situation is analoguous to programmable state discrimination. One difference, however, is that the quantum state 
to which we apply the unitary transforms may be entangled, leading to a richer variety of possible strategies.
By suitable selection of an input state and generalized measurement of the output state, both unambiguous and minimum-error discrimination can be achieved. Pairwise comparison of operators, comparing each transform to be identified with a program transform, is often a useful strategy. There  are, however, situations in which more complicated strategies perform better. This is the case especially when the number of allowed applications of program operations is different from the number of the transforms to be identified.
\bigskip

\end{abstract}

\section{Introduction}

The problem of discriminating between operators in quantum mechanics is closely
related to that of discriminating between quantum states~\cite{Croke}.  In order to
distinguish between operators, a reference state is transformed by the different
operators, and then measurements are performed on the result, so that one is
ultimately distinguishing between the possible output states. Discriminating
between two \emph{known} operators has already been treated~\cite{childs, acin,
  dariano1, chefles1, chefles2}.  Reconstructing a single unknown operator from
measurements on a series of states that have been transformed by it is known as
operator, channel, or quantum process tomography~\cite{dariano2}.  A final
problem, operator comparison, which determines whether two unknown unitary
operators are the same or different~\cite{andersson, andersson2}, is closely related to the
problems considered here.
 
Recently state discrimination problems have been considered in which not all of
the states are known~\cite{bergou1, hayashi1, hayashi2, bergou2}. One may know
some of the states, but is only provided with examples of the others.  For
example, one is given a particle that is guaranteed to be in one of the states
$|\psi_{1}\rangle$ or $|\psi_{2}\rangle$.  The states $|\psi_{1}\rangle$ and
$|\psi_{2}\rangle$ are unknown, but instead, an example of each of these states
is provided.  The object is to use the reference states to best determine
whether the given particle in the unknown state is either $|\psi_{1}\rangle$ or
$|\psi_{2}\rangle$.  Another scenario is that $|\psi_{1}\rangle$ is known, but
$|\psi_{2}\rangle$ is represented by a reference particle, and one is presented
with a particle guaranteed to be in either $|\psi_{1}\rangle$ or
$|\psi_{2}\rangle$. Again the task is to determine the state of the latter
unknown particle.

In this paper we address problems in the realm of operator discrimination,
which are the natural analogues of those, decribed above, for quantum
state discrimination. Instead of having a physical description of the operations, they
are given as actual operations (boxes) which we are allowed to apply to
input states. The situation is reminiscent of oracle quantum computation~\cite{NC},
in which a key component is a device that carries out one of a number of possible
transformation.  Generally, we would like to decide which particular unlabelled
boxes match/correspond to labelled reference boxes or in which order the
unlabelled boxes are placed. The problem can be viewed as programmable pattern
matching, ``programmable'' because we do not know the actual operations as they
are only supplied as reference ``programs''.

We shall consider both minimum error, and unambiguous discrimination tasks. In
minimum-error discrimination, the object is to always make a guess and to
minimize the probability of making a mistake.  In unambiguous discrimination,
the procedure is allowed to fail sometimes (that is, we are allowed sometimes to
refuse to make a guess) but we never want to make a mistake if we do decide to
make a determination. The object is to minimize the probability of failure.

We wish to examine whether good procedures for these decision problems can be
constructed from pairwise operator comparison. The task is to compare two 
unknown operators in order to determine whether or not they are
the same.  This problem was considered in \cite{andersson}, and the best 
procedure found there was to use a singlet state~\cite{andersson}. Each
operator is applied to one of the two qubits in the singlet and the result is
then measured to see if it has a component in the symmetric subspace, upon which
the operators must have been different. We would like to see if this procedure
can be used as a basic unit for more complicated operator comparisons as
discussed above. We shall see that there are cases where singlet-based pairwise
comparison strategies are not optimal.

\section{Pattern matching for unitary operators}

The general scenario which we consider is as follows: Alice/Bob are given
reference boxes, labelled either $U$ or $V$, which implement unitary operations
on single qubits.  The forms of these two unitary operators are not known to
any of the parties. Charlie is given a number of boxes which are unlabelled, but 
each is guaranteed to perform either $U$ or $V$. Our task is to determine what 
boxes Charlie possesses
by applying these boxes to a multi-qubit reference state and then measuring the
results. We shall assume that each box can only be applied once and that $U$ and
$V$ are equally and independently likely to be any qubit unitary. A variety of
scenarios is possible. depending on which boxes are available for Alice, Bob and Charlie.

\begin{figure}
\includegraphics[width=\textwidth]{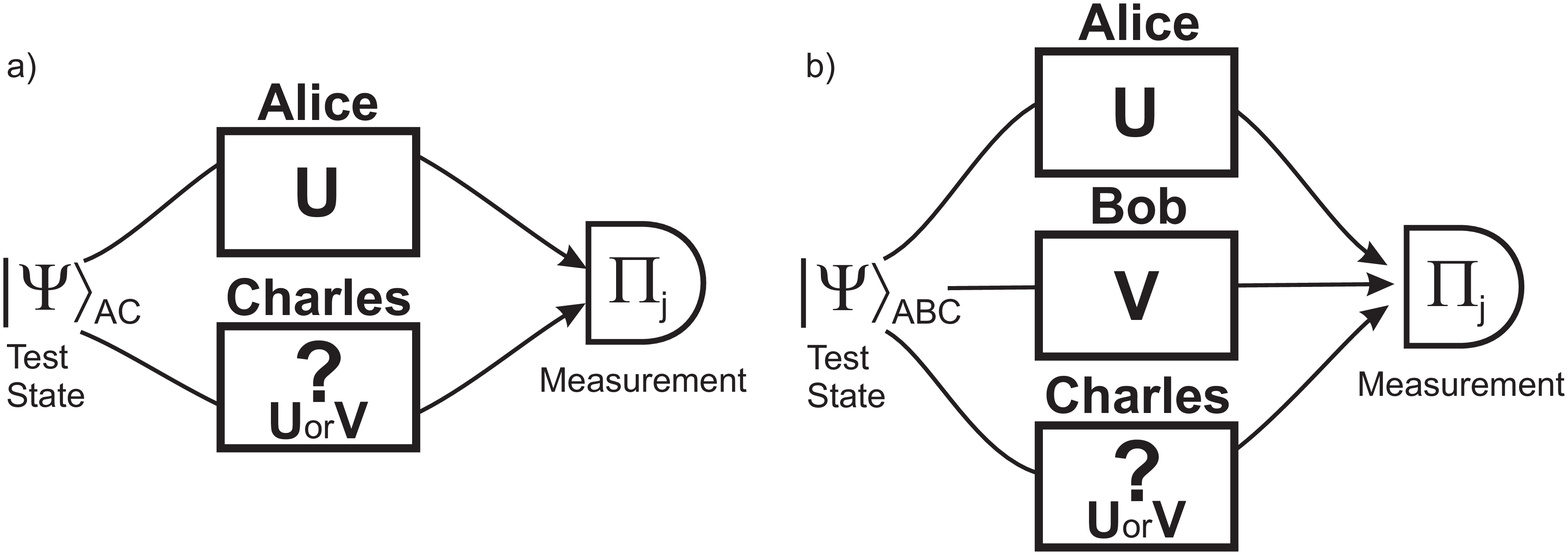}
\caption{One unlabelled box. a) Alice has a reference box $U$, and Charlie has an
  unlabelled box which either performs $U$ or a different unknown unitary $V$. b)
  Alice and Bob each have a different reference box implementing $U$ and $V$
  respectively, and Charlie either has a box which performs $U$ or $V$. We
  prepare a test state $|\Psi\rangle$ which is acted upon by the operators
  and is finally measured leading to an outcome $\Pi_j$.}
\label{fig:OneUnknown}
\end{figure}

\subsection{One reference box and one unlabelled box}

We begin with the simplest case of one reference box and one unlabelled box
which may of may not match the reference (Fig.~\ref{fig:OneUnknown}a). Alice has a
box that performs an unknown operation $U$ and Charlie has one that either
performs $U$ or an unknown operator $V$.
This is similar to operator comparison~\cite{andersson} where it was found that
making use of the two-qubit singlet state, $|\psi^-\rangle_{AB} =
(|0\rangle_{A}|1\rangle_{B} - |0\rangle_{A} |1\rangle_{B})/\sqrt{2}$, by sending
one qubit through each box is a good strategy.  After the qubits have been acted
upon by the boxes, we measure the result to see if it lies in the symmetric or
antisymmetric subspace. If the state is found to be symmetric, then the two
unitary transforms must have been different.

This is in fact the optimal strategy also for the situation we are now
considering, if we confine ourselves to two qubits and demand an unambiguous
result. We have to apply both transforms to a test state $|\Psi\rangle$, giving two possible cases
\begin{equation}
  |\Psi_1\rangle_{AC} = U\otimes U |\Psi\rangle_{AC} ~~
{\rm and}~~|\Psi_2\rangle_{AC} = U\otimes V |\Psi\rangle_{AC} .
\end{equation}
The forms of the single-qubit operators, $U$ and $V$, are unknown, and so we must 
average over $U$ and $V$ to obtain the possible output states
\begin{equation}
\rho_1 = \int du  |\Psi_1\rangle_{AC}\langle\Psi_1 | {\rm ~~and~~}
\rho_2 = \int du dv |\Psi_2\rangle_{AC}\langle\Psi_2|,
\end{equation}
between which we should distinguish. 
The integrals are Haar integrals over $SU(2)$, and can be performed using the following parameterization.  We first set $U=\exp (-i\theta \hat{\mathbf{e}}\cdot \vec\sigma/2)$, 
where $\hat{\mathbf{e}}$ is a unit vector equal to $(\sin\mu \cos\phi ,\sin\mu \sin\phi , \cos\mu )$, with $0\leq \theta \leq 2\pi$, $0\leq \phi \leq 2\pi$, and $0\leq \mu \leq \pi$, and $\vec\sigma=\{\sigma_x,\sigma_y\sigma_z\}$ is a vector of the usual Pauli matrices. The second operator, $V$, is parametrized similarly.  We then have that
\begin{equation}
du=
\frac{1}{4\pi^{2}}\sin^{2}\left(\frac{\theta}{2}\right)\sin\mu d\theta d\phi d\mu .
\end{equation} 
A general 2-qubit test state can be written
\begin{equation}
  |\Psi\rangle_{AC}=
a|\psi^-\rangle + b|\psi^+\rangle + c|\phi^-\rangle + d|\phi^+\rangle ,
\end{equation}
where $|\psi^\pm\rangle=(|01\rangle\pm|10\rangle)/\sqrt{2}$ and
$|\phi^\pm\rangle=(|00\rangle\pm|11\rangle)/\sqrt{2}$ are the usual Bell
states, and $|a|^2+|b|^2+|c|^2+|d|^2 = 1$.  Performing the integrals for this
initial test state, see Appendix \ref{integrals}, we find that
\begin{eqnarray}
\label{isak}
  \rho_1 &=& |a|^2 P^a + \frac{1}{3}(|b|^2+|c|^2+|d|^2)P^s\nonumber\\
  \rho_2 &=& \frac{1}{4}I_A\otimes I_C.
\label{rho12}
\end{eqnarray}
Here $P^a$ and $P^s$ are the projectors onto the antisymmetric and symmetric
subspaces respectively. As the support of $\rho_2$ is all of the two-qubit
space, $\rho_1$ can never be unambiguously identified, as its support will
always be contained in that of $\rho_2$. To unambiguously identify the state
$\rho_2$, the initial test state should lie entirely either in the antisymmetric
or in the symmetric subspace, and we look for a component of the final state in
the symmetric or antisymmetric subspace, respectively.

It is straightforward to check that starting with a singlet state by choosing
$|a|=1$ will give a higher success probability than starting in the symmetric
subspace. The positive operator valued measure (POVM) elements for the optimal
measurement are the projections $\Pi_0 = P^a$, corresponding to the inconclusive
result (failure), and $\Pi_{2}= P^s$, corresponding to identifying the state as
$\rho_2$.  Therefore, assuming that $\rho_1$ and $\rho_2$ are
equally probable, the probability of the measurement failing is
\begin{equation}
p_{f}=\frac{1}{2} [ {\rm Tr}(\Pi_0\rho_1) + {\rm Tr}(\Pi_0\rho_2)] = \frac{5}{8} ,
\end{equation}
so that the success probability is $p_s=3/8$.  An initial test state in the
symmetric subspace would give a success probability of $p_s=1/8$.

Now let us consider the minimum-error strategy.  The optimal
minimum-error measurement for distinguishing between two states $\rho_1$ and
$\rho_2$, with prior probabilities $p_1$ and $p_2$, has the error
probability~\cite{Helstrom}
\begin{equation}
p_e = \frac{1}{2}\left[ 1-{\rm Tr}\sqrt{(p_1\rho_1-p_2\rho_2)^2}\right] ,
\end{equation}
where we shall consider the case $p_1=p_2=1/2$.  Substituting in the
density matrices from Eq.\ (\ref{isak}), we find that
\begin{equation}
{\rm Tr}\sqrt{(\rho_1- \rho_2)^2} = 2\left| \frac{1}{4} - |a|^{2}\right| .
\end{equation}
This is clearly maximized, and the error probability minimized, when $|a|^2 = 1$,
which occurs when the input state is a singlet.  In that case we have that
$\Pi_{1}=P^a$ corresponds to Bob's box being $U$, and $\Pi_{2}=P^s$
corresponds to Bob's box being $V$, and the probability of error is $1/8$.
For the singlet test state, this is the optimal measurement, as the optimal
minimum-error measurement. 

Note that the unambiguous discrimination strategy only gives us an answer when
Charlie's unknown box is $V$.  If Alice instead has a reference box $V$, we
can get an answer only when Bob's box is $U$.  The measurement and success
probability stays the same.

\subsection{Two reference boxes and one unlabelled box}
\label{two1}

Let us now consider a decision problem where Alice has a reference box $U$, Bob
has reference box $V$, and Charlie has an unlabelled box guaranteed to be either
$U$ or $V$ (Fig.~\ref{fig:OneUnknown}b). First let us look at a three-qubit
test state $|\Psi\rangle_{ABC}$ and apply the boxes held by Alice, Bob and
Charlie to their respective qubits.  A general pure three-qubit state
can be written as
\begin{equation}
|\Psi\rangle_{ABC} = a_0|000\rangle + a_1|001\rangle + \ldots + a_7 |111\rangle,
\end{equation}
with $\sum |a_i|^2 = 1$. This means that we are trying to discriminate between
the states
\begin{eqnarray}
\label{3qubits-states}
|\Psi_1\rangle_{ABC} &=& U\otimes V\otimes U |\Psi\rangle_{ABC}\label{three1}\\
|\Psi_2\rangle_{ABC} &=& U\otimes V\otimes V |\Psi\rangle_{ABC}\label{three2}.
\end{eqnarray}
Averaging over $U$ and $V$ means that we have to distinguish between the states
\begin{eqnarray}
\rho_1 &=& \int du dv U\otimes V \otimes U|\Psi\rangle_{ABC}\langle\Psi | U^\dagger\otimes V^\dagger\otimes U^\dagger\nonumber\\
\rho_2 &=& \int du dv U\otimes V \otimes V|\Psi\rangle_{ABC}\langle\Psi | U^\dagger\otimes V^\dagger\otimes V^\dagger .
\end{eqnarray}
Again using the results in Appendix A, we get
\begin{eqnarray}
\rho_1 &=& \frac{1}{4} P^a_{AC}\otimes I_B (|a_1-a_4|^2+|a_3-a_6|^2)\nonumber\\
&&+\frac{1}{6}P^s_{AC}\otimes I_B(|a_0|^2+|a_2|^2+|a_1+a_4|^2+|a_3+a_6|^2+|a_5|^2+|a_7|^2)
\nonumber\\
\rho_2 &=& \frac{1}{4} P^a_{BC}\otimes I_A (|a_1-a_2|^2+|a_5-a_6|^2)\\
&&+\frac{1}{6}P^s_{BC}\otimes I_A(|a_0|^2+|a_3|^2+|a_1+a_2|^2+|a_5+a_6|^2+|a_4|^2+|a_7|^2).\nonumber
\end{eqnarray}
For unambiguous discrimination to be possible, the support of either $\rho_1$ or
$\rho_2$ or both has to be less than all of the three-qubit space. Without loss
of generality, let us assume that $\rho_1$ has support in either
$P^a_{AC}\otimes I_B$ or $P^s_{AC}\otimes I_B$, but not in both. The first
possibility clearly corresponds to a test state which is a singlet in qubits $A$
and $C$, tensored with any state for qubit $B$. This strategy has an overall
success probability of $3/8$. The optimal measurement tests whether or not we have
a singlet in qubits $A$ and $C$, and if not, this unambiguously identifies the state
as $\rho_2$. We could also start with a singlet state in qubits $B$ and $C$, and
test whether we still have a singlet in these qubits after application of the
unitary transforms.

The other possibility is that $\rho_1$ has support in the space symmetric in
qubits $A$ and $C$. In this case $\rho_2$ either has support in all of Hilbert
space, or only in the space symmetric in qubits $B$ and $C$. The latter alternative
gives a higher success probability, as we then have a nonzero probability to
unambiguously identify both $\rho_1$ and $\rho_2$. We then have
\begin{equation}
\rho_1 =  \frac{1}{6} P^s_{AC}\otimes I_B {\rm ~~and~~ }
\rho_2 =  \frac{1}{6} P^s_{BC}\otimes I_A.
\end{equation}
The POVM that unambiguously discriminates between these two density matrices must have
the element that detects $\rho_{1}$ be proportional to $P^a_{BC}\otimes I_A$, because the
only vectors orthogonal to $\rho_{2}$ are of the form $|\psi^{-}\rangle_{BC}\otimes
|\eta\rangle_{A}$, where $|\eta\rangle_{A}$ is an arbitrary state in the space of qubit $A$, 
and by symmetry, this POVM element should treat all vectors in  the space
of qubit $A$ equally.  Similar considerations apply to the element that detects $\rho_{2}$.
The only remaining task is to determine the constant of proportionality that is allowed by the
requirement that the third POVM element, which correspond to the failure of the measurement,
be positive.  Doing so, we find that the optimal measurement in this case has measurement operators
\cite{bergou1,bergou2}
\begin{eqnarray}
\Pi_1 &=& \frac{2}{3}P^a_{BC}\otimes I_A\nonumber\\
\Pi_2 &=& \frac{2}{3}P^a_{AC}\otimes I_B\nonumber\\
\Pi_? &=& I-\Pi_1-\Pi_2,
\end{eqnarray}
and the overall success probability is $1/6$. This is higher than $1/8$, which is
the result if we choose to randomly test for a singlet either in qubits $AC$ or
qubits $BC$. But it is still much lower than $3/8$, which we achieve if we start
with a singlet state either in qubits $AC$ or qubits $BC$.

For the minimum-error task, we can also try to use a state that approximates a
singlet state in qubits $AC$ and $BC$ as well as possible, without obeying the
strict symmetry rules necessary for the unambiguous discrimination strategy.
The state
\[
|\Psi\rangle_{ABC} = \frac{1}{2\sqrt{3}}( |100\rangle + |010\rangle + |011\rangle + |101\rangle ) 
- \frac{1}{\sqrt{3}} ( |001\rangle + |110\rangle )
\]
has the property that its reduced density operators $\rho_{AC}$ and $\rho_{BC}$
have overlap with the singlet state
\[
\langle \psi^-|\rho_{AC}|\psi^-\rangle = \langle \psi^-|\rho_{BC}|\psi^-\rangle = 3/4
\]
where $|\psi^-\rangle$ is a singlet. It is straightforward to prove that the
fidelity of $3/4$ is the best possible if one demands that the fidelities in qubit
pairs $AC$ and $BC$ should be equal.  
After Alice, Bob and Charlie have applied their boxes to obtain (\ref{three1})
and (\ref{three2}), and then averaging over $U$ and $V$, the two
possible output states are
\[
\rho_{1} = \left(\frac{3}{8}P^{a}_{AC} + \frac{1}{24}P^{s}_{AC}\right)\otimes I_{B} ,
\qquad \rho_{2} = \left(\frac{3}{8}P^{a}_{BC} + \frac{1}{24}P^{s}_{BC}\right)\otimes
I_{A},
\]
which then have to be distinguished. The optimum minimum error (Helstrom)
strategy for distinguishing between two density operators $\rho_1$ and $\rho_2$
has the error probability
\[
P_e = \frac{1}{2}\left\{ 1-{\rm Tr}\left[\left( p_1\rho_1 -p_2\rho_2\right)
    ^2\right]^{1/2} \right\} = \frac{1}{2}\left(1-\sum_i|\lambda_i|\right),
\]
where $\lambda_i$ are the eigenvalues of $p_1\rho_1 -p_2\rho_2$. In
our case $p_1=p_2=1/2$, and
\begin{eqnarray}
p_1\rho_1 -p_2\rho_2 &=& 
\frac{1}{2} \left[ \left(\frac{3}{8}P^{a}_{AC} + \frac{1}{24}P^{s}_{AC}\right)\otimes I_{B} -
\left(\frac{3}{8}P^{a}_{BC} + \frac{1}{24}P^{s}_{bC}\right)\otimes I_{A} \right]\nonumber\\
&=& \frac{1}{2} \left[ \left(\frac{1}{3}P^{a}_{AC} + \frac{1}{24}I_{AC}\right)\otimes I_{B} -
\left(\frac{1}{3}P^{a}_{BC} + \frac{1}{24}I_{BC}\right)\otimes I_{A} \right]\nonumber\\
&=& \frac{1}{6}\left[ P^a_{AC}\otimes I_B - P^a_{BC}\otimes I_A\right] \nonumber\\
&=& \frac{1}{12} \left( |100\rangle\langle 100 | +|011\rangle\langle 011 | 
- |010\rangle\langle 010 | - |101\rangle\langle 101 | \right.\nonumber\\
&&\left. -|100\rangle\langle 001 | -|110\rangle\langle 011 | 
+ |010\rangle\langle 001 | + |110\rangle\langle 101 | + h.c. \right) .\nonumber
\end{eqnarray}
This operator is six-dimensional, but happens to be block diagonal with two
blocks of size 3$\times$3. This is easily seen since the off-diagonal elements
only couple states with an equal number of 0's and 1's. Ordering the basis
vectors within the blocks $\left\{ |010\rangle, |100\rangle,
  |001\rangle\right\}$ and $\left\{ |101\rangle, |011\rangle,
  |110\rangle\right\}$, each of the blocks takes the form
\[
\frac{1}{12}\left(\begin{array}{c c c}
-1 & 0 & 1\\
0 & 1 & -1 \\
1 & -1 & 0
\end{array}\right).
\]
The blocks have eigenvalues $\pm 1/(4\sqrt{3})$ and zero, and the sum of the
absolute values of all eigenvalues is therefore $1/\sqrt{3}$. This gives an
error probability of $P_e =(1-1/\sqrt{3})/2 \approx 0.211325$
which is higher than 1/8. Therefore, the simple singlet strategy (a singlet
either in pair $AB$ or $AC$) does better.

\subsubsection{Doing better than pairwise operator comparison}
\label{nonpairwise}

So far, the best strategies we have found are based on pairwise operator
comparison.  Can we do better?  It turns out that if we use a four-qubit state then
we can.  In particular, we shall show that it is possible
to improve the success probabilities for unambiguous discrimination.  We first
note that we can get distinct signals for both operators in the case of
unambiguous discrimination if we use four qubits. Let
\begin{equation}
\label{4qubit}
|\Psi\rangle_{ABCD} = \frac{1}{\sqrt{2}} (|\psi^-\rangle_{AC} |0\rangle_{B} |0\rangle_{D}
+|\psi^-\rangle_{BC} |0\rangle_{A} |1\rangle_{D} ) .
\end{equation}
As usual, Alice, Bob and Charlie apply their respective boxes but no operation
is applied to the fourth qubit D.  It is best to express the resulting states as
density matrices.  If Charlie's box performs $U$, the operator $U\otimes
V\otimes U\otimes I$ is applied to $|\Psi\rangle_{ABCD}$, and after averaging
over both $U$ and $V$ we obtain
\begin{eqnarray}
\label{3boxrho1}
\rho_{1} & = &\frac{1}{4} P^{a}_{AC}\otimes I_{B}\otimes \left(|0\rangle_{D}+\frac{1}{2}|1\rangle_{D}\right)\left
(\,_{D}\langle 0| + \frac{1}{2}\,_{D}\langle 1|\right)  \nonumber \\
 & & +\frac{1}{16}P_{AC}^{s}\otimes I_{B} \otimes |1\rangle_{D}\langle 1| .
\end{eqnarray}
If Charlie's box performs $V$, the operator $U\otimes V\otimes V\otimes I$ is
applied to $|\Psi\rangle_{ABCD}$, and after averaging over both $U$ and $V$ we
obtain
\begin{eqnarray}
\label{3boxrho2}
\rho_{2} & = &\frac{1}{4} P^{a}_{BC}\otimes I_{A}\otimes \left(|1\rangle_{D}+\frac{1}{2}|0\rangle_{D}\right)
\left(\,_{D}\langle 1| + \frac{1}{2}\,_{D}\langle 0|\right)  \nonumber \\
 & & +\frac{1}{16}P_{BC}^{s}\otimes I_{A} \otimes |0\rangle_{D}\langle 0| .
\end{eqnarray}
We will consider only the case of unambiguous discrimination and first choose the POVM
elements to be
\begin{eqnarray}
\label{povm4qubit}
\Pi_{1} & = & P^{s}_{BC}\otimes I_{A}\otimes |1\rangle_{D}\langle 1|  \nonumber \\
\Pi_{2} & = & P^{s}_{AC}\otimes I_{B}\otimes |0\rangle_{D}\langle 0|  \nonumber \\
\Pi_{0} & = & I - \Pi_{1}- \Pi_{2}  =
P^{a}_{BC}\otimes I_{A} \otimes |1\rangle_{D}\langle 1|
+ P^{a}_{AC}\otimes I_{B} \otimes |0\rangle_{D}\langle 0| . \label{three3}
\end{eqnarray}
The element $\Pi_{1}$ corresponds to Charlie's box being $U$, $\Pi_{2}$
corresponds to it being $V$, and $\Pi_{0}$ corresponds to the indeterminate
result.  We find that ${\rm Tr}(\Pi_{1}\rho_{2}) = {\rm Tr}(\Pi_{2}\rho_{1}) =
0$, which guarantees that the discrimination is unambiguous.  In addition, we
have that ${\rm Tr}(\Pi_{1}\rho_{1}) = {\rm Tr}(\Pi_{2}\rho_{2}) = 3/8$, which,
assuming that the boxes for $U$ and $V$ are equally probable, gives a success
probability of $3/8$.  This is the same success probability as with the simpler
two-qubit procedure, but we now have a probability of unambiguously identifying
both boxes, and not just one.

So far, in terms of the overall success probability, we have not gained
anything.  We could in fact also have achieved the same result using a two-qubit
strategy, by choosing to compare Charlie's box against either Alice's $U$ or
Bob's $V$ with probability 1/2 each. The fourth qubit above acts like a quantum
coin. By using the four-qubit test state in (\ref{4qubit}) we can, 
however, do better than the success probability of $3/8$. Though this test state is
most probably not optimal, we can try to improve the POVM.  We make use of the
results of subspace discrimination to find the optimal POVM for the
discrimination of the ranges of the two possible density matrices
\cite{bergou3,rudolph}.  While this does not give the optimal POVM for
discriminating the two density matrices, it does give one that is better than
(\ref{three3}) based on two-operator comparison.  The details of the POVM that
optimally discriminates the ranges of $\rho_{1}$ and $\rho_{2}$ are given in
Appendix B.  One finds that each density matrix can be detected unambiguously
with a probability of $\sim 0.43$, which is greater than $3/8=0.375$.
Therefore, we can conclude that there are better procedures than simple pairwise
operator comparison for solving our three-box decision problem. The optimal
strategy remains to be determined.

\section{Two reference boxes and two unlabelled boxes}
\label{twounknowncopies}

\begin{figure}
\includegraphics[width=\textwidth]{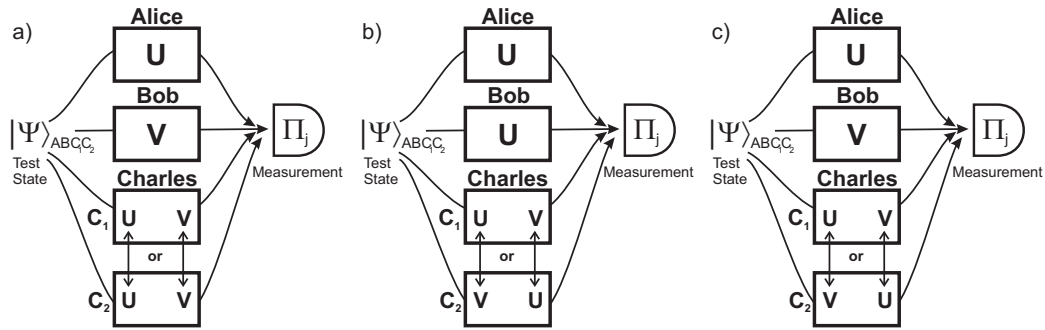}
\caption{Two references and two unlabelled boxes. a) b) c).}
\label{fig:FourBoxes}
\end{figure}

When Charlie possesses two unlabelled boxes, there are several
scenarios we can consider (Fig. \ref{fig:FourBoxes}). Alice and Bob can possess
$U$ and $V$ respectively, Charlie is given two boxes either both $U$ or both
$V$, and the task is to determine which boxes Charlie has. We can also consider
the case in which Alice and Bob possess $U$ and $V$ respectively and Charlie is
given two unlabelled boxes, say $C_1$ and $C_2$, guaranteed that either $C_1=U$
and $C_2=V$ or vice versa, and our task is to determine the order in which $U$
and $V$ appear. For this task, we will also consider the case where the reference
boxes that Alice and Bob are given are either both $U$ or both $V$.

\subsection{Two reference boxes and two of the same unlabelled box}
\label{twotwosame}

Let us consider the case of Alice having box $U$, Bob box $V$, and Charlie
a pair of identical boxes, both of which are $U$ or $V$ (Fig. \ref{fig:FourBoxes}a). 
A good strategy here is to use four qubits in the form of two singlet states,
$|\Psi\rangle_{ABC_1 C_2} = |\psi^-\rangle_{AC_1}|\psi^-\rangle_{BC_2}$ and
apply the boxes.  If both boxes $C_1$  and $C_2$ are $V$, then the output state is
\begin{equation}
  |\Psi_{1}\rangle_{ABC_1 C_2} = (U\otimes V)|\psi^-\rangle_{AC_1}\ (V\otimes V)|\psi^-\rangle_{BC_2} ,
\label{fourqubitstate}
\end{equation}
and if both $C_1$ and $C_2$ are $U$, then we have
\begin{equation}
  |\Psi_{2}\rangle_{ABC_1 C_2} = (U\otimes U)|\psi^-\rangle_{AC_1}\ (V\otimes U)|\psi^-\rangle_{BC_2}.
\label{fourqubitstate2}
\end{equation}
If we now average over both $U$ and $V$, we obtain the possible output density
operators
\begin{eqnarray}
\label{4qubitrho}
\rho_{1} & = & \frac{1}{4} I_{AC_1} \otimes P^{a}_{BC_2}    \nonumber  \\
\rho_{2} & = & \frac{1}{4} P^{a}_{AC_1} \otimes I_{BC_2}  .
\end{eqnarray}
between which we must distinguish.

Let us first consider unambiguous discrimination.  The best POVM elements are
\begin{eqnarray}
\label{fourqubitPOVM}
\Pi_{1} & = & P^{s}_{AC_1}\otimes P^{a}_{BC_2}  \nonumber \\
\Pi_{2} & = &  P^{a}_{AC_1}\otimes P^{s}_{BC_2}   \nonumber  \\
\Pi_{0} & = & P^{s}_{AC_1}\otimes P^{s}_{BC_2} + P^{a}_{AC_1}\otimes P^{a}_{BC_2} ,
\end{eqnarray}
where $\Pi_{0}$ is associated with the failure outcome.  We find that, as is
necessary, ${\rm Tr}(\Pi_{1}\rho_{2}) = {\rm Tr}(\Pi_{2}\rho_{1}) = 0$, and that
${\rm Tr}(\Pi_{1}\rho_{1}) = {\rm Tr}(\Pi_{2}\rho_{2}) = 3/4$.  With both
situations being equally likely, our overall success probability is then $3/4$.

For minimum-error discrimination, we always need a result; the measurement is
not allowed to fail.  This means that we have to do something with the operators
appearing in $\Pi_{0}$.  We first note that the outcome corresponding to $
P^{s}_{AC_1}\otimes P^{s}_{BC_2}$ never occurs, so it can be added to either
$\Pi_{1}$ or $\Pi_{2}$ without any effect.  The outcome corresponding to $
P^{a}_{AC_1}\otimes P^{a}_{B_C2}$ can occur, and for either density matrix does so
with a probability $1/4$.  When it does, we have no idea which density matrix we
have, so we have to guess.  One way of guessing is that we can always guess
$\rho_{2}$, and we shall be wrong half the time.  This corresponds to adding
$P^{a}_{AC_1}\otimes P^{(2)}_{BC_2}$ to $\Pi_{2}$.  (For equal prior probabilities
of $\rho_1$ and $\rho_2$, it does not matter how we guess, but if the
probabilities are unequal, we should always guess the more likely one.)
Therefore, for minimum-error discrimination we can choose for our POVM operators
\begin{eqnarray}
\Pi_{1} & = & P^{s}_{AC_1}\otimes P^{a}_{BC_2} + P^{s}_{AC_1}\otimes P^{s}_{BC_2} \nonumber \\
\Pi_{2} & = &  P^{a}_{AC_1}\otimes P^{s}_{BC_2} + P^{a}_{AC_1}\otimes P^{a}_{BC_2} ,
\end{eqnarray}
and the probability of error will be $1/8$.

\subsection{Two reference boxes and two different unlabelled boxes}

We now turn our attention to a variant of our decision problem in which the object is
to determine which box corresponds to which operator. Charlie has two boxes
$C_1$ and $C_2$, and either box $C_1$ performs $U$ and box $C_2$ performs $V$,
or vice versa. Our object is to determine which of these two situations
holds. This version of the problem is sometimes equivalent to the version with
one unknown box considered in previous sections. For example, if in addition to
Charlie's boxes, there is only one program box $U$, we now have to distinguish
between $U\otimes V\otimes U$ and $V\otimes U\otimes U$. As we are averaging
over $U$ and $V$, we can however interchange these in the second density matrix,
so that $V\otimes U \otimes U$ becomes $U\otimes V\otimes V$. This scenario is
equivalent, therefore, to the third box being unlabelled, with one reference box
each of $U$ and $V$. All the strategies we have discussed above that apply to
this three-box situation can therefore be carried over to the order-finding
version. It remains, therefore, to consider the four-box situations, in which Alice and Bob
both have reference boxes..

\subsubsection{Two reference boxes of $U$}

In the case in which Alice and Bob both have the same reference box, say $U$, and
Charlie has either $C_1=U$ and $C_2=V$, or vice versa
((Fig. \ref{fig:FourBoxes}b)), we can map this case directly onto the case in
Section~\ref{twotwosame}. We note that the two possibilities for $ABC_1 C_2$
are either $U\otimes U\otimes U\otimes V$ or $U\otimes U\otimes V\otimes
U$. As we have to average over all possible $U$ and $V$, this is equivalent
to trying to distinguish between  $U\otimes V\otimes U\otimes U$
and $U\otimes V\otimes V\otimes V$ where we have simply re-ordered the boxes ($C_1 C_2 AB$) and
interchanged $U$ and $V$ in the second case.

\subsubsection{One reference box of $U$ and one reference box of $V$}
\label{orderprogramsUV}

One might wonder if replacing one of the $U$ operators by a $V$ in this scheme
would improve the results.  That is, suppose that Bob has a box that performs
$V$ instead of $U$ (Fig. \ref{fig:FourBoxes}c).  If box $C_1$ is $U$ and box
$C_2$ is $V$, then we now have the state
\begin{equation}
|\Psi_{1}\rangle_{ABC_1C_2} = (U\otimes U)|\psi^-\rangle_{AC_1}\ (V\otimes V)|\psi^-\rangle_{BC_2} ,
\end{equation}
and if box $C_1$ is $V$ and box $C_2$ is $U$, we have
\begin{equation}
|\Psi_{2}\rangle_{ABC_1C_2} = (U\otimes V)|\psi^-\rangle_{AC_1}\ (V\otimes U)|\psi^-\rangle_{BC_2} .
\end{equation}
We average these two states over $U$ and $V$ giving us the two density matrices
\begin{eqnarray}
\rho_{1} & = & |\Psi_1\rangle_{ABC_1C_2}\langle \Psi_1 |   \nonumber  \\
\rho_{2} & = & \frac{1}{4} ( \frac{1}{3} P^{s}_{BC_1}\otimes P^{s}_{AC_2} + P^{a}_{BC_1}\otimes
P^{a}_{AC_2} ) .
\end{eqnarray}
In comparing these two density matrices, it is useful to express $|\Psi\rangle_{ABC_1C_2}$ as
\begin{equation}
|\Psi\rangle_{ABC_1C_2} = \frac{1}{2} ( |\psi^+\rangle_{BC_1}|\psi^+\rangle_{AC_2} +
|\psi^-\rangle_{BC_1}|\psi^-\rangle_{AC_2} - |00\rangle_{BC_1} |11\rangle_{AC_2} - |11\rangle_{BC_1}
|00\rangle_{AC_2} ) .
\end{equation}
The first thing to notice is that the support of $\rho_{1}$ is contained in the
support of $\rho_{2}$, so that if we are considering unambiguous discrimination,
we will be able to positively identify $\rho_{2}$, but we will never be able to
positively identify $\rho_{1}$.

We choose $\Pi_{2}$, the POVM element corresponding to $\rho_{2}$, to be the sum
of three projection operators.  The first is the projection onto the part of
$\mathcal{H}^{s}_{BC_1}\otimes \mathcal{H}^{s}_{AC_2}$ that is orthogonal to the
vector
\begin{equation}
|q\rangle =  |\psi^+\rangle_{BC_1}|\psi^+\rangle_{AC_2}  - 
|00\rangle_{BC_1} |11\rangle_{AC_2}
 - |11\rangle_{BC_1} |00\rangle_{AC_2} ) ,
\end{equation}
the second is the projection onto $\mathcal{H}^{s}_{BC_1}\otimes
\mathcal{H}^{a}_{AC_2}$, and the third is the projection onto
$\mathcal{H}^{a}_{BC_1}\otimes \mathcal{H}^{s}_{AC_2}$.  The notation here is
that $\mathcal{H}^{s}_{jk}$ is the symmetric subspace of qubits $j$ and $k$,
while $\mathcal{H}^{a}_{jk}$ is the antisymmetric subspace of qubits $j$ and
$k$.  The POVM element corresponding to failure is $\Pi_{0} = I - \Pi_{2}$.  We
find that ${\rm Tr}(\Pi_{2}\rho_{2})=2/3$, which means the our overall success
rate is $1/3$; we succeed with a probability of $2/3$ if we have $\rho_{2}$,
which happens with a probability of $1/2$, and we get an indeterminate answer if
we are given $\rho_{1}$.

We have found that replacing one of the $U$ operators by $V$ has made the situation
worse. This makes sense as detecting a difference in either the pair $AC_1$ or
$BC_2$ will identify the order of $U$ and $V$ for boxes $C_1$ and $C_2$. If we
match up identical operations in both singlet pairs, then we have no chance of
unambiguously identifying the order of boxes $C_1$ and $C_2$. Matching different
transforms up in both pairs gives a higher probability to detect a difference,
as opposed to when there is a difference in only one singlet pair. This cannot,
however, make up for our inability to detect a difference when identical
transforms were matched up. It may be that in this situation, a strategy with
more than four qubits would help. This is because the situation is somewhat
similar to the three-box situation in Section~\ref{nonpairwise}. There, we would
like to compare the unlabelled box against a reference box which does not match
it, in order to unambiguously identify the unlabelled box. We cannot know which
reference box to choose, however, and half the time we will choose to compare
the unlabelled box against an identical reference box. The four-qubit strategy
somehow helps us beat the odds. Here, we would like to compare both unlabelled
boxes with reference boxes which do not match, but half the time we make the
wrong choice and match the unlabelled boxes up with matching reference boxes. A
strategy with more qubits might do better, but this is still an open problem.

\section{Conclusion}

We have discussed a number of operator decision problems.  In each we are set the 
task of determining whether boxes perform the same or different unitary transformations,
$U$ or $V$.  We have no prior information of the forms of the transformations, but, 
instead, we have reference boxes that are known to perform the transformations $U$ or $V$.

We began by considering one reference box and one unknown box.  This situation reduced
to operator comparison, and for both unambiguous discrimination and minimum-error discrimination
it is best to send into each box one of the particles from a singlet state.  We then went on to examine
the situation where one has two reference boxes and one unknown box.  This can be thought of
as a kind of pattern matching.  Each of the reference operators represents a pattern, and we are 
trying to determine which of the two patterns the unknown operator matches.  We studied
both unambiguous and minimum-error strategies based on pairwise operator comparison.  We 
also studied a minimum-error strategy based on an entangled three-qubit state that best approximates
two singlet states.  This strategy was not as good as pairwise operator comparison.  Using
subspace discrimination, we found an unambiguous strategy that is better than pairwise operator
comparison, which demonstrates that the pairwise comparison strategy is not optimal.  Finally,
we studied situations with two reference boxes and two unknown boxes.  For these cases, we
only studied strategies based on pairwise operator comparison.

Our objective was not to present a comprehensive theory of operator comparison problems, but to
to gain some idea of the issues involved and to present some strategies that will work for 
specific problems.  One conclusion is that pairwise operator comparison works well, but is
not always optimal.  Finding optimal strategies is a subject that deserves further study.

\vskip1cm

\noindent {\bf Acknowledgments}

\noindent DKLO acknowledges the support of the Scottish Universities Physics Alliance
(SUPA) and the Quantum Information Scotland Network (QUISCO). MH acknowledges support from the Royal Society of Edinburgh and from the SUPA Distinguished Visitor programme.  SMB thanks the Royal Society
and the Wolfson Foundation for their generous support.

\vskip1cm

\appendix

\section{}

\label{integrals}

Here we present some additional information on the Haar integrals that
were used extensively throughout the paper.  As was noted, for $SU(2)$ these
integrals can be performed using the following parameterization.  We first set
$U=\exp (-i\theta \hat{\vec{e}}\cdot \vec{\sigma}/2)$, where $\hat{\vec{e}}$ is
a unit vector equal to $(\sin\mu \cos\phi ,\sin\mu \sin\phi , \cos\mu )$, with
$0\leq \theta \leq 2\pi$, $0\leq \phi \leq 2\pi$, and $0\leq \mu \leq \pi$, and
$\vec{\sigma}=\{\sigma_x,\sigma_y,\sigma_z\}$ is a vector of the usual Pauli
matrices. We then have that
\begin{equation}
du =
\frac{1}{4\pi^{2}}\sin^{2}\left(\frac{\theta}{2}\right)\sin\mu d\theta d\phi d\mu .
\end{equation} 
By explicit integration we find that
\begin{eqnarray}
\int du U|0\rangle\langle 0|U^{\dagger} & = & \int du U|1\rangle\langle 1|U^{\dagger} = \frac{1}{2} I   ,  \nonumber  \\
\int du U|0\rangle\langle 1| U^{\dagger}& = & \int du U|1\rangle\langle 0|U^{\dagger} = 0.
\end{eqnarray}
The unitary invariance of the measure then implies that for any qubit state
$|\psi\rangle$
\begin{eqnarray}
\int du U|\psi\rangle\langle \psi|U^{\dagger} & = & \frac{1}{2} I   ,  \nonumber \\
\int du U|\psi \rangle\langle \psi_{\perp}| U^{\dagger} & = & 0,
\end{eqnarray}
where $\langle\psi |\psi_{\perp}\rangle = 0$.  

Now let us move on to two qubits.  One can explicitly evaluate the integrals
\begin{eqnarray}
  \int du U\otimes U |00\rangle\langle 00| U^{\dagger}\otimes U^{\dagger}& = & \frac{1}{3}P^{s} \nonumber \\
  \int du U\otimes U |00\rangle\langle 11| U^{\dagger}\otimes U^{\dagger} & = & 0  \nonumber \\
  \int du U\otimes U |\psi^-\rangle\langle 00| U^{\dagger}\otimes U^{\dagger} & = & 0 .
\end{eqnarray} 
The unitary invariance of the measure allows us to immediately extend these
results
\begin{eqnarray}
  \int du U\otimes U |\psi_{s}\rangle\langle \psi_{s}| U^{\dagger}\otimes U^{\dagger}& = & \frac{1}{3}P^{s} 
  \nonumber \\
  \int du U\otimes U |\psi_{s}\rangle\langle \psi_{s\perp}| U^{\dagger}\otimes U^{\dagger} & = & 0  
  \nonumber \\
  \int du U\otimes U |\psi^-\rangle\langle \psi_{s} | U^{\dagger}\otimes U^{\dagger} & = & 0 ,
\end{eqnarray} 
where $|\psi_{s}\rangle$ and $|\psi_{s\perp}\rangle$ are in the symmetric
subspace, $\langle\psi_{s}|\psi_{s\perp}\rangle =0$, and $|\psi_{a}\rangle$ is
in the antisymmetric subspace.

\section{}

Here we present the details of the POVM that optimally discriminates between the
ranges of $\rho_{1}$ and $\rho_{2}$ in Eqs. (\ref{3boxrho1}) and
(\ref{3boxrho2}).  Let $S_{1}$ denote the range of $\rho_{1}$.  It is spanned by
the orthonormal vectors
\begin{eqnarray}
  |u_{1}\rangle = \frac{2}{\sqrt{5}}|\psi^-\rangle_{AC}|0\rangle_{B}\left( |0\rangle_{D}
    +\frac{1}{2}|1\rangle_{D}\right), &\qquad& |u_{5}\rangle = |11\rangle_{AC}|0\rangle_{B}|1\rangle_{D}
  \nonumber \\ 
  |u_{2}\rangle = \frac{2}{\sqrt{5}}|\psi^-\rangle_{AC}|1\rangle_{B}\left( |0\rangle_{D}
    +\frac{1}{2}|1\rangle_{D}\right), &\qquad& |u_{6}\rangle = |11\rangle_{AC}|1\rangle_{B}|1\rangle_{D}
  \nonumber \\ 
  |u_{3}\rangle = |00\rangle_{AC}|0\rangle_{B}|1\rangle_{D}, &\qquad& |u_{7}\rangle = |\psi^+\rangle_{AC}
  |0\rangle_{B}|1\rangle_{D} \nonumber \\
  |u_{4}\rangle =  |00\rangle_{AC}|1\rangle_{B}|1\rangle_{D}, &\qquad& |u_{8}\rangle = |\psi^+\rangle_{AC}|1\rangle_{B}|1\rangle_{D}  .
\end{eqnarray}
Here, $|\psi^+\rangle = (|0\rangle |1\rangle + |1\rangle |0\rangle )/\sqrt{2}$.
In this basis, we have
\begin{equation}
\rho_{1}  =  \frac{5}{16}( |u_{1}\rangle\langle u_{1}| + |u_{2}\rangle\langle u_{2}| )
+ \frac{1}{16}\sum_{j=3}^{8} |u_{j}\rangle\langle u_{j}|  .
\end{equation}
Now, let $S_{2}$ denote the range of $\rho_{2}$.  It is spanned by the
orthonormal vectors
\begin{eqnarray}
  |v_{1}\rangle = \frac{2}{\sqrt{5}}|\psi^-\rangle_{BC}|0\rangle_{A}\left( |1\rangle_{D}
    +\frac{1}{2}|0\rangle_{D}\right), &\qquad& |v_{5}\rangle = |11\rangle_{BC}|0\rangle_{A}|0\rangle_{D}
  \nonumber \\ 
  |v_{2}\rangle = \frac{2}{\sqrt{5}}|\psi^-\rangle_{BC}|1\rangle_{A}\left( |1\rangle_{D}
    +\frac{1}{2}|0\rangle_{D}\right), &\qquad& |v_{6}\rangle = |11\rangle_{BC}|1\rangle_{A}|0\rangle_{D}
  \nonumber \\ 
  |v_{3}\rangle = |00\rangle_{BC}|0\rangle_{A}|0\rangle_{D}, &\qquad& |v_{7}\rangle = |\psi^+\rangle_{BC}
  |0\rangle_{A}|0\rangle_{D} \nonumber \\
  |v_{4}\rangle =  |00\rangle_{BC}|1\rangle_{A}|0\rangle_{D}, &\qquad& |v_{8}\rangle = |\psi^+\rangle_{BC}|1\rangle_{A}|0\rangle_{D}  .
\end{eqnarray}
In this basis
\begin{equation}
  \rho_{2}  = \frac{5}{16} ( |v_{1}\rangle\langle v_{1}| + |v_{2}\rangle\langle v_{2}|) + \frac{1}{16}
  \sum_{j=3}^{8} |v_{j}\rangle\langle v_{j}|  .
\end{equation}
Examining these vectors, we see that $|u_{3}\rangle$ and $|u_{6}\rangle$ are
orthogonal to $S_{2}$, and $|v_{3}\rangle$ and $|v_{4}\rangle$ are orthogonal to
$S_{1}$.  The remaining vectors can be arranged into two sets of overlapping
subspaces.  The subspaces spanned by $\{ |u_{1}\rangle , |u_{4}\rangle ,
|u_{7}\rangle \}$ and by $\{ |v_{1}\rangle , |v_{4}\rangle , |v_{7}\rangle \}$
overlap and the subspaces spanned by $\{ |u_{2}\rangle , |u_{5}\rangle ,
|u_{8}\rangle \}$ and $\{ |v_{2}\rangle , |v_{5}\rangle , |v_{8}\rangle \}$
overlap.  The subspaces in the first set are orthogonal to those in the second.

We can now apply the results of reference \cite{bergou3} to discriminate the two
three-dimensional subspaces and to discriminate the two two-dimensional
subspaces.  These results are then combined with those of the previous paragraph
to give us the two optimal POVM's for unambiguously discrimination $S_{1}$ and
$S_{2}$.  In order to exhibit these POVM's it is necessary to define a number of
additional vectors.  We have in $S_{1}$
\begin{eqnarray}
|\tilde{u}_{1}\rangle & = & \frac{1}{\sqrt{3}}(|u_{4}\rangle + \sqrt{2}|u_{7}\rangle ) \nonumber \\
|\tilde{u}_{2}\rangle & = & \frac{1}{\sqrt{3}}(|u_{5}\rangle +\sqrt{2} |u_{8}\rangle )
\end{eqnarray}
and in $S_{2}$
\begin{eqnarray}
|\tilde{v}_{1}\rangle & = & \frac{1}{\sqrt{3}}(|v_{4}\rangle +\sqrt{2} |v_{7}\rangle ) \nonumber \\
|\tilde{v}_{2}\rangle & = & \frac{1}{\sqrt{3}}(|v_{5}\rangle + \sqrt{2}|v_{8}\rangle ) .
\end{eqnarray}
Both $|\tilde{u}_{1}\rangle$ and $|\tilde{u}_{2}\rangle$ are constructed so that
they are orthogonal to $S_{2}$ and $|\tilde{v}_{1}\rangle$ and
$|\tilde{v}_{2}\rangle$ are orthogonal to $S_{1}$.  Our discrimination problem
is now reduced to discriminating between two sets of two-dimensional subspaces.
The first set consists of the subspaces spanned by $\{ |u_{1}\rangle ,
|\tilde{u}_{3}\rangle \}$ and $\{ |v_{1}\rangle , |\tilde{v}_{3}\rangle \}$, and
the second set consists of the subspaces spanned by $\{ |u_{2}\rangle ,
|\tilde{u}_{4}\rangle \}$ and $\{ |v_{2}\rangle , |\tilde{v}_{4}\rangle \}$,
where
\begin{eqnarray}
|\tilde{u}_{3}\rangle = \frac{1}{\sqrt{3}}(\sqrt{2}|u_{4}\rangle -|u_{7}\rangle )&\qquad &
|\tilde{u}_{4}\rangle = \frac{1}{\sqrt{3}}(\sqrt{2}|u_{5}\rangle - |u_{8}\rangle ) \nonumber \\
|\tilde{v}_{3}\rangle = \frac{1}{\sqrt{3}}(\sqrt{2}|v_{4}\rangle - |v_{7}\rangle )&\qquad &
|\tilde{v}_{4}\rangle = \frac{1}{\sqrt{3}}(\sqrt{2}|v_{5}\rangle -|v_{8}\rangle ) .
\end{eqnarray}
The two sets are orthogonal to each other.  The next step is to find the Jordan
bases for each set of subspaces \cite{bergou3}.  In each case, we find that the
two subspaces have a vector in common.  Throwing these vectors out, because the
clearly do not help discriminate the subspaces, we are left with two sets of one
dimensional subspaces to discriminate; $|\tilde{u}_{5}\rangle$ and
$|\tilde{v}_{5}\rangle$ on the one hand, and $|\tilde{u}_{6}\rangle$ and
$|\tilde{v}_{6}\rangle$ on the other.  These vectors are defined by
\begin{eqnarray}
|\tilde{u}_{5}\rangle = \frac{1}{2\sqrt{2}}(\sqrt{3}|u_{1}-\sqrt{5}|\tilde{u}_{3}\rangle )&\qquad &
|\tilde{u}_{6}\rangle = \frac{1}{2\sqrt{2}}(\sqrt{3}|u_{2}+\sqrt{5}|\tilde{u}_{4}\rangle ) \nonumber \\
|\tilde{v}_{5}\rangle = \frac{1}{2\sqrt{2}}(\sqrt{3}|v_{1}-\sqrt{5}|\tilde{v}_{3}\rangle )&\qquad &
|\tilde{v}_{6}\rangle = \frac{1}{2\sqrt{2}}(\sqrt{3}|v_{2}+\sqrt{5}|\tilde{v}_{4}\rangle ) .
\end{eqnarray}
Our final step is to define the vectors
\begin{eqnarray}
|y_{5}\rangle = \frac{5}{4}(|\tilde{v}_{5}\rangle +\frac{3}{5}|\tilde{u}_{5}\rangle ), &\qquad &
|z_{5}\rangle  =  \frac{5}{4} (|\tilde{u}_{5}\rangle +\frac{3}{5}|\tilde{v}_{5}\rangle ) \nonumber \\
|y_{6}\rangle  =  \frac{5}{4}(|\tilde{v}_{6}\rangle +\frac{3}{5}|\tilde{u}_{6}\rangle )&\qquad&
|z_{6}\rangle  =  \frac{5}{4} (|\tilde{u}_{6}\rangle +\frac{3}{5}|\tilde{v}_{6}\rangle ) .
\end{eqnarray}
These vectors have the property that $\langle y_{j}|\tilde{u}_{j}\rangle = 0$
and $\langle z_{j}|\tilde{v}_{j}\rangle = 0$. for $j=5,6$.  The POVM operators
for optimally unambiguously distinguishing $S_{1}$ and $S_{2}$, and thereby
distinguishing $\rho_{1}$ and $\rho_{2}$, are
\begin{eqnarray}
\Pi_{1} & = & |u_{3}\rangle\langle u_{3}| + |u_{6}\rangle\langle u_{6}| 
+ |\tilde{u}_{1}\rangle \langle\tilde{u}_{1}| + |\tilde{u}_{2}\rangle\langle\tilde{u}_{2}| \nonumber \\
& & + \frac{5}{8}(|z_{5}\rangle \langle z_{5}| + |z_{6}\rangle \langle z_{6}|)  \nonumber \\
\Pi_{2} & = & |v_{3}\rangle\langle v_{3}| + |v_{6}\rangle\langle v_{6}| 
+ |\tilde{v}_{1}\rangle \langle\tilde{v}_{1}| + |\tilde{v}_{2}\rangle\langle\tilde{v}_{2}| \nonumber \\
& & + \frac{5}{8}(|y_{5}\rangle \langle y_{5}| + |y_{6}\rangle \langle y_{6}|)  
\end{eqnarray}

\end{document}